\documentclass[
  aps,
  pra,
  reprint,
  amsmath,
  amssymb,
  floatfix,
  superscriptaddress
]{revtex4-2}

\usepackage[dvipdfmx]{graphicx}
\usepackage{bm}
\usepackage{color}

\providecommand{\ket}[1]{\left|#1\right\rangle}
\providecommand{\bra}[1]{\left\langle#1\right|}

\DeclareMathOperator{\Tr}{Tr}
\newcommand{\Rey}{\operatorname{Re}}

\begin{document}

\title{Limits of independent and identical measurements for quantum
illumination with an unknown return phase}

\author{Ko Shiraiwa}
\affiliation{National Defense Academy of Japan, Yokosuka, Japan}

\author{Shingo Kukita}
\email{kukita@nda.ac.jp}
\affiliation{National Defense Academy of Japan, Yokosuka, Japan}

\begin{abstract}
Quantum illumination exploits entanglement between a transmitted signal and a retained idler to improve the error-probability exponent of target detection by roughly a factor of four (6 dB) over that with a coherent state of the same transmitted energy. This advantage presumes a known return phase. In practice, the phase is set by the range to the target and the condition of its surface, and is difficult to know in advance. Whether the advantage survives when this phase is unknown is not obvious. Here, we cast target detection as a composite hypothesis test in which the return phase is an unknown constant common to all trials, and we restrict the receiver to independent and identical measurements on each copy. We bound the worst-case error exponent at low reflectivity for every such measurement and every input state of a single signal mode and an idler of any dimension. We then show that unentangled coherent light with heterodyne detection already saturates this bound, for every value of the phase. Entanglement therefore confers no advantage in this setting. The class of independent and identical measurements contains many implementable quantum-illumination receivers, including the optical parametric amplifier and phase-conjugate receivers. Our result shows that none of them can offer a quantum advantage in the worst case over the phase, to leading order in the reflectivity.
\end{abstract}

\maketitle

\section{Introduction}
\label{sec:intro}

The detection of a target under strong background noise and large propagation
loss is a basic problem in radar and lidar. Quantum illumination (QI) is a
framework in which the transmitted signal beam is entangled with an idler
beam~\cite{Lloyd2008,Tan2008}. The returned light is then measured jointly with
the idler retained at the receiver. Under strong background noise, QI achieves
an error-probability exponent, given by the quantum Chernoff information,
roughly four times ($6$~dB) that of detection with a coherent state of the same
transmitted energy~\cite{Tan2008}. QI has also been
demonstrated experimentally, in both the optical and the microwave
regimes~\cite{Zhang2013,Zhang2015,Barzanjeh2020}.

The choice of the input state on the transmitter side has been studied
extensively, and the two-mode squeezed vacuum (TMSV) repeatedly emerges as an
optimal or near-optimal choice under a range of criteria and
constraints~\cite{DePalma2018,Bradshaw2021,NairGu2020}.
The design of the receiver has likewise received much attention. Measurements that address each
copy individually cannot in general attain the quantum Chernoff information~\cite{Calsamiglia2008}. The optical parametric amplifier (OPA) and phase-conjugate receivers belong to this class; they are easy to implement, but the error exponent they achieve is only twice ($3$~dB) that with a coherent state~\cite{GuhaErkmen2009}.
Receivers that attain the full factor of four include the sum-frequency-generation (SFG) receiver and its
feed-forward variant~\cite{Zhuang2017PRL}, as well as the more easily
implementable correlation-to-displacement receiver~\cite{Shi2022,Shi2024}.
These receivers act across
the copies rather than addressing each one individually.

All of these results presume that the phase of the light returned from the
target is known. The OPA and phase-conjugate receivers, for instance, read out
the phase-sensitive correlation between the return mode and the idler mode, and
this readout requires a phase reference. In practice, however, the return phase
depends on the round-trip distance to the target and on the condition of its
surface, and cannot in general be predicted in advance.

The effect of an unknown phase has so far been examined for specific receivers.
For a fading target, whose return amplitude is Rayleigh distributed and whose
return phase is uniformly distributed, the signal-to-noise ratio of the OPA
receiver falls below that of detection with a coherent state~\cite{Zhuang2017}.
The same work shows that the SFG receiver can nevertheless cope with such a
random phase. Results for the correlation-to-displacement receiver under the
same fading model have also been reported~\cite{Chen2023}.
As noted above, both receivers act across the copies.

Here we ask how well independent and identical measurements (IIM), in which the
same POVM acts on each copy, can perform when the return phase is unknown. We
take the return phase $\theta$ to be an unknown constant common to all trials,
as is the case when the target moves slowly compared with the measurement time.
Target detection is then a composite hypothesis test with the unknown phase
parameter $\theta$. In this setting, we derive an upper bound on the worst-case
error exponent at low reflectivity, for any IIM and any input state of a single
signal mode and an idler of arbitrary dimension. We then show that this bound is
saturated, for every value of $\theta$, by an entanglement-free configuration: a
coherent state with heterodyne detection.
To leading order in the reflectivity, the quantum advantage therefore disappears.

The IIM class contains many implementable receivers, including the OPA and
phase-conjugate receivers. Our result implies that none of them can deliver a
quantum advantage when the return phase is unknown. It follows that any
receiver that does retain an advantage must either adapt its measurement across
the copies or process them jointly, as the SFG and correlation-to-displacement
receivers do~\cite{Zhuang2017,Chen2023}.

The paper is organized as follows. Section~\ref{sec:setup} summarizes the
formulation of quantum illumination and prior work. Section~\ref{sec:bound}
derives the upper bound for IIM with an unknown phase and shows that it is
attained even with an unentangled state. Section~\ref{sec:discussion} discusses
the results.

\section{Formulation of quantum illumination and prior work}
\label{sec:setup}

We formulate target detection as a binary hypothesis test. Let $H_{0}$ denote
the situation in which the target is absent and $H_{1}$ the situation in which
it is present. A transmitter radiates light toward the region to be
interrogated, and a receiver detects the returned light. Under $H_{0}$, the
receiver collects only thermal radiation from the background; under $H_{1}$, it
collects a mixture of thermal radiation and the transmitted light. This process
is repeated $M$ times, and the two hypotheses are discriminated from the
measurement outcomes. When the thermal radiation is much stronger than the
signal, the two hypotheses become difficult to distinguish.

\begin{figure}[t]
    \centering
    \includegraphics[width=\columnwidth]{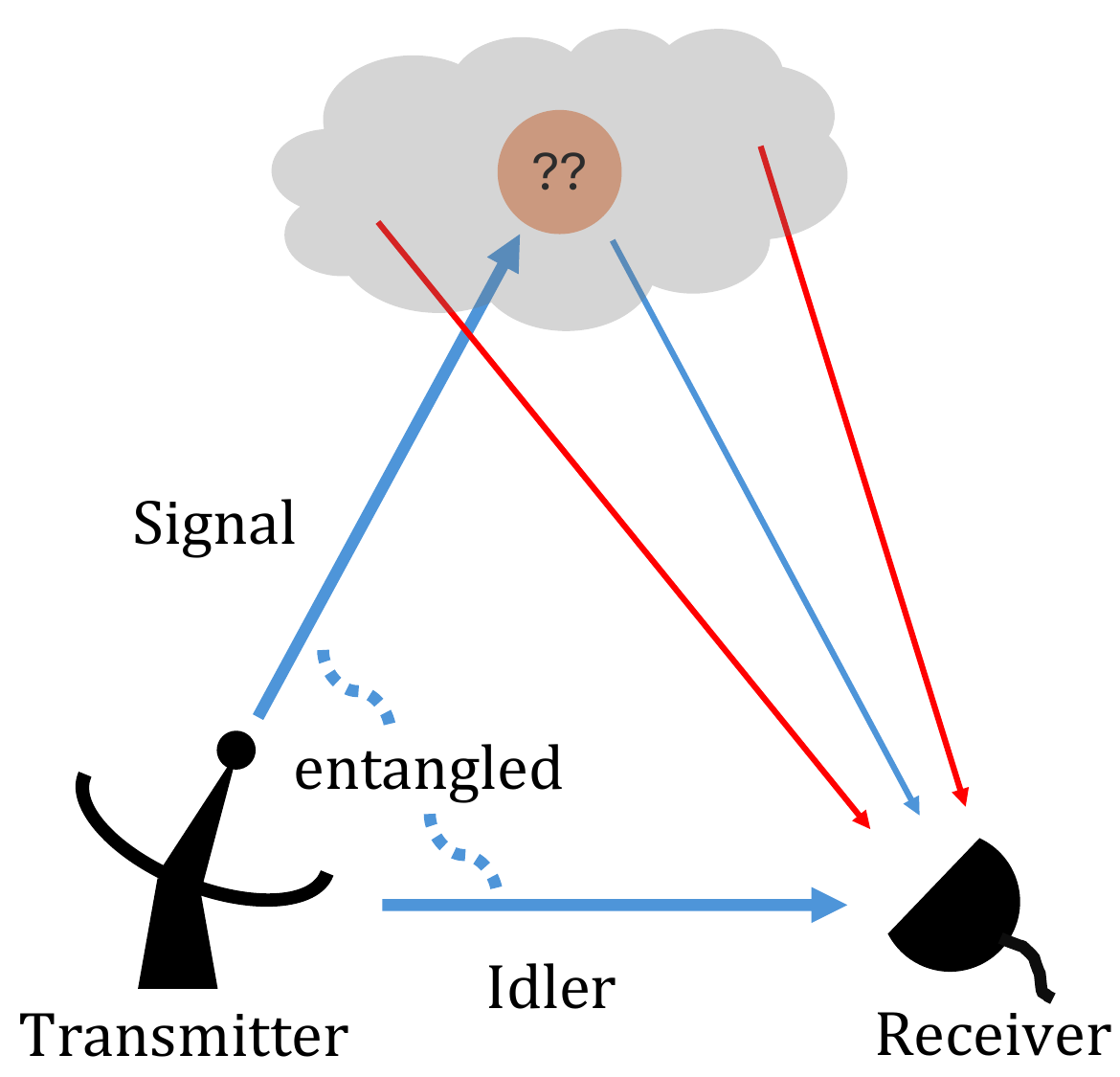}
    \caption{Schematic of quantum illumination. The transmitter prepares an entangled state of a signal and an idler, radiates the signal toward the region to be interrogated, and passes the idler to the receiver. If a target is present, part of the signal is reflected back to the receiver (blue). The receiver also collects thermal radiation from the background (red) in either case. Measuring the returned light together with the idler, the receiver decides whether a target is present.}
    \label{fig:protocol}
\end{figure}

In quantum illumination, the transmitted light and an idler are prepared in an
entangled state, and the idler is retained at the receiver
(Fig.~\ref{fig:protocol}). In what follows, we write $S$ for the signal system
and $I$ for the idler, and assume that the signal consists of a single bosonic
mode. The signal Hilbert space $\mathcal{H}_{S}$ is then the Fock space spanned
by $\{\ket{n}_{S}\}_{n=0,1,2,\dots}$, with annihilation operator $\hat{a}_{S}$.
The idler Hilbert space $\mathcal{H}_{I}$ is left arbitrary.

We prepare an entangled pure state $\ket{\psi}_{SI}$ of the two systems; for
example, a two-mode squeezed vacuum (TMSV) is often used in quantum
illumination. The target is idealized as a beam splitter of reflectivity
$\kappa$ that mixes the signal mode with the environment, another single bosonic
mode $E$, whose annihilation operator we write as $\hat a_{E}$. The beam-splitter
unitary is
\begin{equation}
    \hat U=\exp\left[-i\varphi\left(\hat a_S^\dagger\hat a_E
    +\hat a_S\hat a_E^\dagger\right)\right],
    \qquad \sin\varphi=\sqrt\kappa .
    \label{eq:uni_ideal}
\end{equation}
Before mixing, the environment mode is in the thermal state
\begin{equation}
    \rho_E(n)=\sum_{k=0}^{\infty}\frac{n^k}{(n+1)^{k+1}}\ket{k}_{E}\bra{k},
\end{equation}
with mean photon number $n$. After the unitary $\hat U$, mode $S$ now represents
the part of the light that passes the target and never reaches the receiver,
while $E$ becomes the return mode carrying the light reflected back to the
receiver. The receiver uses $E$ and the idler $I$ for the measurement. This
setting also covers entanglement-free schemes that use the return mode alone,
including coherent-state detection: if $\ket{\psi}_{SI}$ is a product state, the
idler plays no role.

Tracing out mode $S$, we obtain the state when the target is present,
\begin{equation}
    \rho_1=\Tr_S\left[\hat U\left(\ket{\psi}_{SI}\bra{\psi}
    \otimes\rho_E\!\left(N_B^{(1)}\right)\right)\hat U^\dagger\right].
\end{equation}
The absence of the target corresponds to $\kappa=0$, and the state is then
\begin{equation}
    \rho_0=\rho_I\otimes\rho_E\!\left(N_B^{(0)}\right),\qquad
    \rho_I=\Tr_S\left[\ket{\psi}_{SI}\bra{\psi}\right].
    \label{eq:hyp0_ideal}
\end{equation}
Here $N_B^{(1)}$ and $N_B^{(0)}$ denote the mean photon number of the
environment mode under $H_{1}$ and $H_{0}$, respectively.

The signal has mean photon number
$N_{S}=\Tr_{S}[\rho_{S}\hat{a}_{S}^{\dagger}\hat{a}_{S}]$, where
$\rho_{S}=\Tr_{I}[\ket{\psi}_{SI}\bra{\psi}]$. We require that $\rho_1$ be
indistinguishable from $\rho_0$ when $N_S=0$, that is, when no signal is
transmitted. For this to hold, the background noise reaching the receiver must
have the same mean photon number $N_B$ under $H_{0}$ and $H_{1}$. In the
beam-splitter model, the mean photon number is $(1-\kappa)N_B^{(1)}$ under
$H_{1}$ and $N_B^{(0)}$ under $H_{0}$. Setting both equal to $N_B$ fixes
\begin{equation}
    N_B^{(0)}=N_B,\qquad N_B^{(1)}=\frac{N_B}{1-\kappa}.
\end{equation}

Repeating the process $M$ times gives $M$ copies of $\rho_0$ under $H_{0}$, and
of $\rho_1$ under $H_{1}$. The minimum error probability $P_e$ then obeys
\begin{equation}
    P_e\leq\frac{1}{2}\exp(-M\xi_Q)
\end{equation}
for any $M$~\cite{Audenaert2007}, where
\begin{equation}
    \xi_Q=-\ln\min_{0\leq s\leq1}\Tr\left[\rho_0^{s}\rho_1^{1-s}\right]
\end{equation}
is the quantum Chernoff information. The bound is attained asymptotically as
$M\to\infty$~\cite{NussbaumSzkola2009}. Thus $\xi_Q$ serves as the figure of
merit for quantum illumination.

Among schemes without entanglement, using a coherent state as the signal is
optimal as $\kappa\to0$~\cite{Bradshaw2021}. In the regime of a weak signal
$N_S\ll1$, a strong background $N_B\gg1$, and low reflectivity $\kappa\ll1$, the
quantum Chernoff information for a coherent state is~\cite{GuhaErkmen2009}
\begin{equation}
    \xi_Q^{\mathrm{coh}}\simeq\frac{\kappa N_S}{4N_B},
    \label{eq:cohe_known}
\end{equation}
whereas for a TMSV it is
\begin{equation}
    \xi_Q^{\mathrm{TMSV}}\simeq\frac{\kappa N_S}{N_B},
\end{equation}
four times as large as $\xi_Q^{\mathrm{coh}}$~\cite{Tan2008}.

The quantum Chernoff information $\xi_Q$ is the exponent obtained after
optimizing over all measurements, including collective measurements on the $M$
copies. Collective measurements are in general difficult to implement.
Independent and identical measurements (IIM), which apply the same POVM to each
copy, are easier, and they make the $M$ outcomes independent and identically
distributed. Within the IIM class, the two hypotheses yield distributions $P_0$
and $P_1$ of the measurement outcome. The minimum error probability then decays
exponentially in $M$, and the exponent is the classical Chernoff information
\begin{equation}
    \xi_C=-\ln\min_{0\leq s\leq1}\sum_x P_0(x)^s P_1(x)^{1-s}.
\end{equation}
In the low-reflectivity regime, $P_0$ and $P_1$ are close to each other, and
$s=1/2$ is then approximately optimal. At $s=1/2$ the Chernoff information
coincides with the Bhattacharyya distance.

For a coherent state, homodyne detection in each trial is an instance of IIM,
and it already achieves $\xi_Q^{\mathrm{coh}}$~\cite{GuhaErkmen2009}. For a
TMSV, the optical parametric amplifier (OPA) receiver and the phase-conjugate
receiver have been proposed as IIM receivers~\cite{GuhaErkmen2009}; for
$\kappa\ll1$ and $N_B\gg1$ they achieve
\begin{equation}
    \xi_C^{\mathrm{OPA}}\simeq\frac{\kappa N_S}{2N_B},
\end{equation}
twice $\xi_Q^{\mathrm{coh}}$ but only half of $\xi_Q^{\mathrm{TMSV}}$. The IIM
class cannot in general attain the quantum Chernoff
information~\cite{Calsamiglia2008}.

The results above all presume a known return phase. For a fading target, whose
return amplitude is Rayleigh distributed and whose return phase is uniformly
distributed, the error probability of any receiver decays only
subexponentially~\cite{Zhuang2017}. A quantum advantage nevertheless survives
for receivers acting across the copies. The sum-frequency-generation (SFG)
receiver, for example, outperforms coherent-state detection under this fading
model~\cite{Zhuang2017}. Whether the IIM class retains a quantum advantage under
an unknown return phase is not obvious.

\section{Error exponent for IIM under an unknown return phase}
\label{sec:bound}

We now let the return phase $\theta$ be unknown. Its effect is the
transformation $\hat{a}_{S}\rightarrow e^{-i\theta}\hat{a}_{S}$, under which the
unitary of Eq.~\eqref{eq:uni_ideal} becomes
\begin{equation}
    \hat U(\theta)=\exp\left[-i\varphi\left(e^{i\theta}\hat a_S^\dagger\hat a_E
    +e^{-i\theta}\hat a_S\hat a_E^\dagger\right)\right].
\end{equation}
The state under $H_{1}$ then depends on the phase,
\begin{equation}
    \rho_{1}(\theta)=\Tr_{S}\left[\hat U(\theta)\left(\ket{\psi}_{SI}\bra{\psi}
    \otimes\rho_{E}(N^{(1)}_{B})\right)\hat U(\theta)^\dagger\right],
\end{equation}
whereas the state under $H_{0}$ is still given by
Eq.~\eqref{eq:hyp0_ideal}.

In the limit of low reflectivity $\kappa\rightarrow0$, we expand
$\rho_1(\theta)$ around $\rho_{0}$, the state at $\kappa=0$:
\begin{equation}
    \rho_1(\theta)=\rho_0+\sqrt\kappa\left(e^{i\theta}\hat V
    +e^{-i\theta}\hat V^\dagger\right)+\kappa\hat\Delta+O(\kappa^{3/2}).
\end{equation}
Here
\begin{align}
    \hat V&=- i \hat{\Lambda}_{I}\otimes\left[\hat a_E,\rho_{E}(N_B)\right],
    \nonumber\\
    \hat{\Lambda}_{I}&=\Tr_S\left[\ket{\psi}_{SI}\bra{\psi}\,
    \hat a_S^\dagger\right].
    \label{eq:V}
\end{align}
Note that $N^{(1)}_{B}=N_B/(1-\kappa)$ is itself expanded in $\kappa$, and
$\rho_E$ in Eq.~\eqref{eq:V} is therefore evaluated at $N_{B}$. Because
$\hat\Delta$ does not contribute to the calculations below, we omit its explicit
form. Trace preservation, $\Tr\rho_1(\theta)=1$ for every $\kappa$, gives
$\Tr[\hat V]=\Tr[\hat\Delta]=0$.

The unknown phase $\theta$ is assumed to take a value common to all trials. The
receiver is restricted to the IIM class. Let $\{\hat{M}_{x}\}_{x}$ denote its
POVM; whether the outcome $x$ is discrete or continuous does not affect the
analysis below. The two hypotheses give
\begin{equation}
    P_{0}(x)=\Tr\left[\rho_{0}\hat{M}_{x}\right],\qquad
    P_{1}(x|\theta)=\Tr\left[\rho_{1}(\theta)\hat{M}_{x}\right].
\end{equation}
Distinguishing them in the presence of the unknown phase $\theta$ is a composite
hypothesis test, whose error exponent is characterized by
\begin{equation}
    \Xi_{C}:=\min_{\theta} \xi_{C}(\theta).
\end{equation}
The minimization reflects the fact that no value of $\theta$ can be excluded in
advance.

We now examine an upper bound on $\Xi_{C}$. We expand $P_{1}(x|\theta)$ in
powers of $\kappa$:
\begin{equation}
    P_{1}(x|\theta)=P_0(x)+2\sqrt\kappa\,\Rey\left[e^{i\theta}v(x)\right]
    +\kappa\,\delta(x)+O(\kappa^{3/2}),
\end{equation}
with $v(x)=\Tr[\hat V\hat{M}_x]$ and $\delta(x)=\Tr[\hat\Delta\hat{M}_x]$. These
quantities satisfy
\begin{equation}
    \sum_{x}v(x)=0,\qquad \sum_{x}\delta(x)=0,
    \label{eq:prob_conserv}
\end{equation}
where $\sum_{x}$ becomes an integral when $x$ is continuous.

Next we expand the Chernoff information $\xi_{C}(\theta)$ to second order in
$\sqrt{\kappa}$:
\begin{align}
    \xi_{C}(\theta)
    &=-\ln\min_{s}\left[\sum_{x}\left(P_{0}(x)\right)^{s}
      \left(P_{1}(x|\theta)\right)^{1-s}\right]\nonumber\\
    &=\frac{\kappa}{2}\sum_{x}
      \frac{\left(\Rey\left[e^{i\theta}v(x)\right]\right)^{2}}{P_{0}(x)}
      +O(\kappa^{3/2}).
\end{align}
Here we used Eq.~\eqref{eq:prob_conserv} and the fact that $s=1/2$ is optimal to
this order. The division by $P_{0}(x)$ in this expression is legitimate for the
following reason. Because $\hat U(\theta)$ does not act on the idler, the
reduced state of $\rho_{1}(\theta)$ on the idler side equals $\rho_{I}$. The
support of a bipartite state is contained in the tensor product of the supports
of its reduced states, and $\rho_{E}(N_{B})$ has full rank; hence
$\mathrm{supp}\,\rho_{1}(\theta)\subseteq\mathrm{supp}\,\rho_{0}$. Therefore
$P_{1}(x|\theta)=0$ whenever $P_{0}(x)=0$, and it suffices to take the sum over
$x$ on the support of $P_{0}$. Separating the phase-dependent and
phase-independent parts of the leading-order term, we obtain
\begin{align}
    \xi_{C}(\theta)=&\frac{\kappa}{4}\sum_{x}\frac{|v(x)|^{2}}{P_{0}(x)}
    \nonumber\\
    &+\frac{\kappa}{4}\Rey\left[e^{2i\theta}\sum_{x}
    \frac{\left(v(x)\right)^{2}}{P_{0}(x)}\right]+O(\kappa^{3/2}).
    \label{eq:phase_unknown}
\end{align}

Minimizing Eq.~\eqref{eq:phase_unknown} over $\theta$ gives
\begin{align}
    \Xi_{C}=&\frac{\kappa}{4}\left(\sum_{x}\frac{|v(x)|^{2}}{P_{0}(x)}
    -\left|\sum_{x}\frac{\left(v(x)\right)^{2}}{P_{0}(x)}\right|\right)
    +O(\kappa^{3/2})\nonumber\\
    \leq&\frac{\kappa}{4}\sum_{x}\frac{|v(x)|^{2}}{P_{0}(x)}+O(\kappa^{3/2}).
    \label{eq:first}
\end{align}
Equality holds when the phase-dependent term in Eq.~\eqref{eq:phase_unknown}
vanishes, that is, when the Chernoff information is independent of $\theta$ to
leading order.

To examine the upper bound on $\Xi_{C}$, we introduce the Schmidt decomposition
of the initial state $\ket{\psi}_{SI}$,
\begin{equation}
    \ket{\psi}_{SI}=\sum_{i}\sqrt{\lambda_{i}}\ket{u_{i}}_{S}\ket{v_{i}}_{I},
    \qquad \lambda_{i}>0 .
    \label{eq:schmit}
\end{equation}
Let ${\cal K}_{S}$ denote the subspace spanned by $\ket{u_{i}}_{S}$, and
${\cal K}_{I}$ the subspace spanned by $\ket{v_{i}}_{I}$; in general
${\cal K}_{S}$ and ${\cal K}_{I}$ are proper subspaces of ${\cal H}_{S}$ and
${\cal H}_{I}$. Introducing the inverse of the density operator
$\rho_I=\Tr_S[\ket{\psi}_{SI}\bra{\psi}]$ on ${\cal K}_{I}$,
\begin{equation}
    \tilde{\rho}^{-1}_{I}=\sum_{i}\frac{\ket{v_{i}}_{I}\bra{v_{i}}}{\lambda_{i}},
    \label{eq:gen_inv}
\end{equation}
we have $\rho_{I}\tilde{\rho}^{-1}_{I}=\tilde{\rho}^{-1}_{I}\rho_{I}
=\hat{\Pi}_{I}$. Note that $\hat{\Pi}_{I}$ is the projector onto ${\cal K}_{I}$
and not the identity operator on the system $I$. On the other hand, the thermal
state $\rho_{E}(N_{B})$ has full rank, and its inverse $\rho^{-1}_{E}$ is
(formally) defined on the whole space of the system $E$,
\begin{equation}
    \rho^{-1}_{E}(N_{B})=\sum^{\infty}_{k=0}
    \frac{(N_{B}+1)^{k+1}}{N_{B}^{k}}\ket{k}_{E}\bra{k},
\end{equation}
satisfying $\rho_{E}\rho^{-1}_{E}=\rho^{-1}_{E}\rho_{E}=\hat{I}_{E}$. We define
the formal inverse of $\rho_{0}=\rho_{I}\otimes\rho_{E}(N_{B})$ by
$\tilde{\rho}^{-1}_{0}=\tilde{\rho}^{-1}_{I}\otimes\rho^{-1}_{E}(N_{B})$.

As is clear from Eq.~\eqref{eq:V}, $\hat{\Lambda}_{I}$ is an operator on
${\cal K}_{I}$, and we may therefore write
\begin{equation}
    v(x)=\Tr\left[\hat{V}\hat{M}_{x}\right]
    =\Tr\left[\left(\hat{\Pi}_{I}\otimes\hat{I}_{E}\right)
    \hat{V}\hat{M}_{x}\right].
\end{equation}
We apply the Cauchy--Schwarz inequality
$|\Tr[\hat{B}^\dagger\hat{A}]|^2\leq\Tr[\hat{A}^\dagger\hat{A}]\,
\Tr[\hat{B}^\dagger\hat{B}]$ to
$\hat{A}=\tilde{\rho}_0^{-1/2}\hat{V}\hat{M}_x^{1/2}$ and
$\hat{B}=\rho_{0}^{1/2}\hat{M}_x^{1/2}$. These operators satisfy
$\Tr[\hat{B}^\dagger\hat{A}]=v(x)$ and $\Tr[\hat{B}^\dagger\hat{B}]=P_0(x)$, and
we obtain
\begin{equation}
    |v(x)|^2\leq P_0(x)\,
    \Tr\left[\hat V^\dagger\tilde{\rho}_0^{-1}\hat V\hat{M}_x\right].
\end{equation}
We divide both sides by $P_0(x)$ and sum over $x$. Using $\sum_x\hat{M}_x=\hat
I$, we obtain
\begin{equation}
    \sum_x\frac{|v(x)|^2}{P_0(x)}
    \leq\Tr\left[\hat V^\dagger\tilde{\rho}_0^{-1}\hat V\right].
    \label{eq:second}
\end{equation}

Using the explicit form of $\hat{V}$ in Eq.~\eqref{eq:V}, we find
\begin{align}
    \Tr\left[\hat V^\dagger\tilde{\rho}_0^{-1}\hat V\right]
    =&\Tr_{I}\left[\hat{\Lambda}^\dagger_{I}\tilde{\rho}_{I}^{-1}
    \hat{\Lambda}_{I}\right]
    \Tr_{E}\left[\frac{\hat{a}^{\dagger}_{E}\rho_{E}\rho^{-1}_{E}
    \rho_{E}\hat{a}_{E}}{(N_{B}+1)^2}\right]\nonumber\\
    =&\frac{\Tr_{I}\left[\hat{\Lambda}^\dagger_{I}\tilde{\rho}_{I}^{-1}
    \hat{\Lambda}_{I}\right]}{N_{B}+1},
    \label{eq:third}
\end{align}
where we used the identity
$[\hat{a}_{E},\rho_{E}(N_{B})]=-\rho_{E}(N_{B})\hat{a}_{E}/(N_{B}+1)$ together
with the canonical commutation relation
$[\hat{a}_{E},\hat{a}^{\dagger}_{E}]=1$. The Schmidt decomposition
[Eq.~\eqref{eq:schmit}] and Eq.~\eqref{eq:gen_inv} then give
\begin{align}
    \Tr_{I}&\left[\hat{\Lambda}^\dagger_{I}\tilde{\rho}_{I}^{-1}
    \hat{\Lambda}_{I}\right]
    =\sum_{i,j}\lambda_{i}\,{}_{S}\bra{u_{i}}\hat{a}^{\dagger}_{S}
    \ket{u_{j}}_{S}\bra{u_{j}}\hat{a}_{S}\ket{u_{i}}_{S}\nonumber\\
    &=\Tr_{S}\left[\rho_{S}\hat{a}_{S}^{\dagger}\hat{\Pi}_{S}
    \hat{a}_{S}\right]
    \leq\Tr_{S}\left[\rho_{S}\hat{a}_{S}^{\dagger}\hat{a}_{S}\right]=N_{S},
    \label{eq:fourth}
\end{align}
where $\hat{\Pi}_{S}$ is the projector onto ${\cal K}_{S}$. The last inequality
follows from $\hat{\Pi}_{S}\leq\hat{I}_{S}$.

Substituting Eqs.~\eqref{eq:second}, \eqref{eq:third}, and \eqref{eq:fourth}
into Eq.~\eqref{eq:first}, we obtain
\begin{equation}
    \Xi_{C}\leq\frac{\kappa N_{S}}{4(N_{B}+1)}+O(\kappa^{3/2}).
    \label{eq:bound}
\end{equation}
This inequality holds for every pure initial state. A mixed input state
$\rho_{SI}$ also reduces to the treatment above: it suffices to take a
purification with a reference system and include that system in the idler.

A quantum advantage would remain if entangled light could saturate
Eq.~\eqref{eq:bound} while unentangled light could not. This is not the case:
unentangled coherent light with heterodyne detection already saturates the
bound, as we confirm in the Appendix. 
In summary, for a single bosonic signal mode, the worst-case exponent in the IIM class is no larger than that of unentangled coherent light.

\section{Summary and discussion}
\label{sec:discussion}

In this work, we asked whether quantum illumination in the IIM class can exhibit
a quantum advantage when the return phase is an unknown constant. When the phase
is known, entangled light in this class gives roughly twice the Chernoff
information of the scheme using unentangled coherent light. We have shown that,
when the phase is unknown, this advantage disappears to leading order in the
reflectivity. More precisely, the worst-case exponent over the phase does not
exceed that of unentangled coherent light. It follows that, under an unknown
return phase, quantum illumination can surpass unentangled detection only if the
measurement is adaptive or collective.

The present work is limited in three respects. We restricted the analysis to
low reflectivity, and the bound in Eq.~\eqref{eq:bound} accordingly holds only
to leading order in $\kappa$. We also addressed only the error exponent, and
evaluating the prefactor is left for future work. Finally, we took the signal
to be a single mode. When the return phase differs from mode to mode, as is the
case for frequency modes, the unknown parameter becomes multidimensional, and
the structure of the worst-case evaluation itself changes. The extension to
multiple modes remains open.

\appendix

\section{Saturation of the bound by a coherent state and heterodyne detection}
\label{sec:coherent}

Consider an initial state in which the signal mode is a coherent state
$\ket{\alpha}_{S}$ with $\alpha\in{\mathbb C}$, so the mean photon number is
$N_S=|\alpha|^2$. This state carries no entanglement with the idler, and the
idler therefore plays no role in what follows. If the target is present, the
reflected signal is mixed with the thermal noise of the environment, and the
return mode becomes a thermal state of mean photon number $N_B$ displaced by an
amplitude $\sqrt\kappa\,e^{i\theta}\alpha$. If the target is absent, the return
mode is an undisplaced thermal state of the same mean photon number.

We discriminate these two states by heterodyne detection, whose POVM is
$\{\ket{\beta}\bra{\beta}/\pi\}_{\beta\in{\mathbb C}}$. The outcome
distribution is a complex Gaussian centered at $0$ under $H_0$ and at
$\sqrt\kappa\,e^{i\theta}\alpha$ under $H_1$,
\begin{align}
    P_0(\beta)=&\frac{1}{\pi(N_B+1)}
    \exp\!\left(-\frac{|\beta|^2}{N_B+1}\right),\nonumber\\
    P_1(\beta|\theta)=&\frac{1}{\pi(N_B+1)}
    \exp\!\left(-\frac{|\beta-\sqrt\kappa\,e^{i\theta}\alpha|^2}{N_B+1}\right).
\end{align}
The two distributions are Gaussian with equal variance and differ only in their
means. The integral defining $\xi_C(\theta)$ can therefore be evaluated in
closed form for every $s$,
\begin{equation}
    \int d^{2}\beta\,P_{0}(\beta)^{s}P_{1}(\beta|\theta)^{1-s}
    =\exp\left[-\frac{s(1-s)\,\kappa|\alpha|^{2}}{N_{B}+1}\right],
\end{equation}
where the minimum over $s$ is attained at $s=1/2$. We thus obtain
\begin{equation}
    \xi_{C}(\theta)=\frac{\kappa N_{S}}{4(N_{B}+1)},
    \label{eq:coh_exact}
\end{equation}
which holds exactly for every $\kappa$ and every $\theta$ and therefore
saturates the bound of Eq.~\eqref{eq:bound} to leading order in $\kappa$.

Finally, we address the attainability of this exponent. In a composite
hypothesis test, an exponent need not be attainable by a receiver ignorant of
the parameter, and the attainability has to be established by construction.
Here such a construction is available. For the $M$ heterodyne outcomes
$\beta_{1},\dots,\beta_{M}$, take the statistic $T=|\sum_{m}\beta_{m}|^{2}$ and
declare the target present when $T/[M(N_{B}+1)]$ exceeds $\gamma/4$, where
$\gamma=M\kappa N_{S}/(N_{B}+1)$. The normalized statistic $T/[M(N_{B}+1)]$ has
unit mean under $H_{0}$ and mean $1+\gamma$ under $H_{1}$, and the error
probabilities of both hypotheses decay as $\exp[-\gamma/4]$. Note that neither
the statistic nor the threshold involves $\theta$, because taking the modulus
of the sum discards its phase. The exponent $\kappa N_{S}/[4(N_{B}+1)]$ is thus
attainable even when the phase is unknown, and so is the bound of
Eq.~\eqref{eq:bound}.

\bibliography{references}

\end{document}